\documentclass[sigconf, nonacm]{acmart}              % arXiv version.
\AtBeginDocument{%
  \providecommand\BibTeX{{%
    \normalfont B\kern-0.5em{\scshape i\kern-0.25em b}\kern-0.8em\TeX}}}

\acmConference[BeyondFacts'24]{4th International Workshop on Computational Methods for Online Discourse Analysis}{May 13--17, 2024}{Singapore}

\setcopyright{acmcopyright}
\copyrightyear{2024}
\acmYear{2024}
\acmDOI{XXXXXXX.XXXXXXX}
\acmPrice{00.00}
\acmISBN{978-1-4503-9419-2/24/05}
\begin{document}
\title{The Web unpacked: a quantitative analysis of global Web usage}

%% Authors:
\author{Henrique S. Xavier}
%\begin{comment}
\email{hxavier@nic.br}
\orcid{0000-0002-9601-601X}
\affiliation{%
  \institution{NIC.br}
  \streetaddress{Av. das Nações Unidas, 11541, 7º andar}
  \city{São Paulo}
  \state{SP}
  \country{Brazil}
  \postcode{04578-000}
}
% Authors for header:
%\renewcommand{\shortauthors}{Xavier et al.}
\renewcommand{\shortauthors}{H. S. Xavier}
%\end{comment}

%% Abstract:
\begin{abstract}

This paper presents a comprehensive analysis of global web usage patterns based on data from SimilarWeb, a leading source for estimating web traffic. Leveraging a dataset comprising over 250,000 websites, we estimate the total web traffic and investigate its distribution among domains and industry sectors. We detail the characteristics of the top 116 domains, which comprise an estimated one-third of all web traffic. Our analysis scrutinizes various attributes of these domains, including their content sources and types, access requirements, offline presence, and ownership features. Our analysis reveals a significant concentration of web traffic, with a diminutive number of top websites capturing the majority of visits. Search engines, news and media, social networks, streaming, and adult content emerge as primary attractors of web traffic, which is also highly concentrated on platforms and USA-owned websites. Much of the traffic goes to for-profit but mostly free-of-charge websites, highlighting the dominance of business models not based on paywalls.
  
\end{abstract}

%\begin{comment}

%% Keywords:

%% A "teaser" image:
%  \includegraphics[width=\textwidth]{images/dalle_smiley_network_teaser}
%  \caption{A painting expressing the associations between sentiments.}
%  \Description{A painting of smiley faces connected by lines as nodes in a network.}

\received{19 April 2024}
%\received[revised]{12 March 2009} % arXiv
%\received[accepted]{5 June 2009}  % arXiv
%\end{comment}

\maketitle

\section{Introduction}
\label{sec:intro}

The World Wide Web (also called the Web) is an online decentralized and owner-free information system released to the public domain in 1993 \cite{BernersLee1990, CERN1993}. It is an application of the Internet protocol suite, a computer communication protocol that is also decentralized and owner-free. The Web itself has no moderation or gatekeepers and, for a long time, had almost no governmental regulation. It was designed to be highly system-independent, technically simple, easy to implement, and permissive to format and content -- all features associated with the Web's massive success and adoption worldwide \cite{Gandon2022}. Currently, the Web is the best-known and most pervasive Internet application. Web browsing is a dominant computer activity \cite{Crichton2021}, and Web-based companies figure among the largest and most valuable companies in the World \cite{Evans2016}. Moreover, HTML, CSS, JavaScript, and WebAssembly technologies have promoted the Web as a universal platform for building computer applications \cite{Bouras2015, Taivalsaari2011}.

The importance of the Web, its lack of explicit management, and its vast potential have raised questions on what are its realized, actual characteristics in terms of main uses, applications, and content types, how visits are distributed among websites, who are the content publishers, how is the content typically produced, and much more.  Answering these questions can provide valuable insights for developing public policies related to the Web, such as regulating new activities, relationships, and business models to promote positive outcomes and reduce negative impacts on society.

In the past, the Web has been thoroughly analyzed in terms of its content and graph properties (e.g., the number of webpages covering a given topic \cite{Chakrabarti2002, Huizingh2000} and the number of hyperlinks connecting them \cite{Adamic2000, Huberman2001}). However, previous accounts about the characteristics of the Web have primarily been based on metrics other than usage (e.g., link structure, number of websites, or companies' market value) or were built on anecdotal accounts and non-quantitative data. Some exceptions are: \cite{Adamic1999, Webster2002} that showed for a restricted subset of users and websites that visitors are distributed among websites in a power-law-like fashion; and \cite{Agarwal2022}, that used Alexa's ranking list to select the 100 most visited domains, Google trends to estimate their popularity and Fortiguard's classification to verify the popularity of different topics, among other things.

This paper aims to provide a data-driven general picture of web usage. Based on data about the number of monthly visits to web domains, we aim to estimate:
\begin{enumerate} 
\item the number of domains (and which ones) required to form a representative picture of global web usage; 
\item the importance of topics explored on the Web; and 
\item the prevalence of broad website characteristics (e.g., access barriers, content sources, and ownership) typically encountered in web visits. 
\end{enumerate}
Our complete analysis can be accessed at \url{http://github.com/cewebbr/web-unpacked}.

This paper is organized as follows: Sec. \ref{sec:data} presents the characteristics of our data and our cleaning process. In Sec. \ref{sec:mom}, we analyze the temporal variation of monthly visits to estimate the stability of our findings. We describe the distribution of visits in Sec. \ref{sec:analysis}, where we also estimate the total monthly visits of the whole Web. We distribute the visits among industries in Sec. \ref{sec:content}. Sec. \ref{sec:top-domains} delves into a detailed analysis of the 116 most visited domains, exploring ownership, content production, login requirements, and the provision of applications. Finally, we summarize our analysis and present our conclusions in Sec. \ref{sec:conclusions}.

\section{Data Set}
\label{sec:data}

The decentralized nature of the Web complicates our ability to understand its characteristics. Without a centralized index and unified access log, accurate measurements of web traffic (such as monthly website visits) and links become challenging \cite{Scheitle2018, Jansen2022}. Regarding web traffic measurement, there are three basic methods:
\begin{enumerate}
\item User-centric: This method involves using software (e.g., browsers, extensions, or other applications) installed on the user's device to monitor that user's behavior.
\item Server-centric: Here, JavaScript and HTTP request logs are employed to monitor accesses and behavior on a given website.
\item Network-centric: This method involves an Internet Service Provider (ISP) monitoring HTTP requests from its customers.
\end{enumerate}
SimilarWeb, a company that combines all three methods, reportedly tracks a representative panel of users to ensure that all domains visited a minimum number of times are monitored \cite{SimilarWeb2024}. Other companies that provide similar data are Semrush and Ahrefs.

The data analyzed in this paper are from SimilarWeb's Starter Plan, covering August to October 2023. This plan provides data about the most visited domains in each of the 210 industry segments specified by SimilarWeb, limited to 10,000 domains per industry. During the covered period, data was available for 1,336,963 domains. For each domain, SimilarWeb estimates the average monthly visits, month-over-month visit variation, average number of unique visitors, average visit duration, and fraction of visits from mobile and desktop. Due to limited accuracy, monthly visits and unique visitors lower than 5,000 figure as ``$<5,000$'' \cite{SimilarWeb2024}. For some low-ranking domains, these data might show as ``-''. In general, subdomains are not distinguished from their parent domains (e.g., visits to docs.google.com are accounted for as part of visits to google.com along with visits to other Google subdomains), and transitions from one subdomain to another do not count as an extra domain visit, just as a single one.

The quality of SimilarWeb data has been evaluated in previous academic works and is considered good enough for relative ballpark measurements between domains. Jansen \emph{et al.} used Google Analytics for 86 websites as a truth table and identified a systematic bias of 20\% in SimilarWeb's monthly visits estimates \cite{Jansen2022}. Prantl compared monthly visits from SimilarWeb and
NetMonitor\footnote{\url{https://www.spir.cz/projekty/netmonitor/}}
for 485 Czech websites and found an average absolute difference of about 42\% between these two sets of measurements \cite{Prantl2018}. Enterprise reports have indicated similar scenarios \cite{Hardwick2018, Diachuk2021}. However, given that our data on monthly visits cover over seven orders of magnitude (from 5 thousand to 86 billion), the reported deviations do not significantly affect the overall distribution of visits among websites. The data still provides precise ranks, with a Pearson correlation of 95\% between monthly visit estimates from SimilarWeb and Google Analytics \cite{Jansen2022} and a Spearman correlation of 96\% between SimilarWeb's and NetMonitor's ranks \cite{Prantl2018}. Also, the present work relies only on relative measurements, so any systematic biases are irrelevant.

It is important to note that web traffic exhibits annual patterns that could impact the representativeness of our three-month data sample \cite{Liu2017,Liu2018}. By analyzing five years of Google Trends data for the most visited domain names across various industries, we found that Western e-commerce sites like amazon.com and walmart.com see increased searches before Christmas, while searches for weather-related websites like weather.com and accuweather.com rise during school vacations. Assuming these annual patterns in Google Trends reflect similar patterns in actual domain visits, we estimate that our data may deviate from its annual average by up to 24\%, as seen in the case of espn.com. However, major domains like youtube.com, instagram.com, and xvideos.com show only minor variations (i.e., less than 5\%). These estimates suggest that, while seasonality is a factor, it is unlikely to affect our findings significantly.    

In our analysis, we excluded domains with values of ``$<5,000$'' and ``-'' for average monthly visits. For each industry, we assumed that SimilarWeb is complete up to a certain threshold (the domain with the lowest average monthly visits). In other words, we assumed that domains missing from an industry's list must have average monthly visits lower than the least visited domain in the list. However, it is essential to note that being complete in a given industry sector does not imply the entire dataset is complete, as SimilarWeb's Starter Plan limits the number of domains per industry to 10,000. To ensure completeness in the entire dataset, we must enforce a threshold on average monthly visits of 140,484, the largest individual industry threshold. This complete dataset, considered in our analysis, contains 254,661 domains and will be ranked by average monthly visits.

\section{Analysis}
\label{sec:analysis}

\subsection{Month-over-month variation}
\label{sec:mom}

\begin{figure}[t]
  \centering
  \includegraphics[width=\columnwidth]{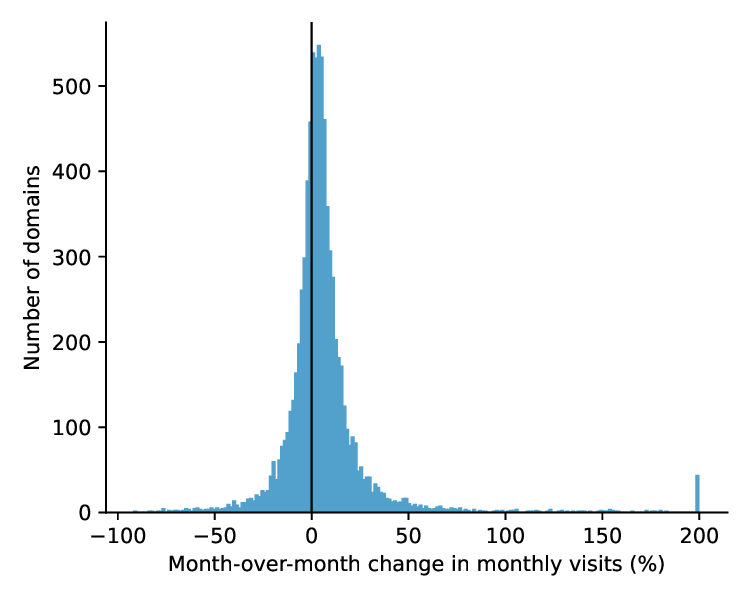}
  \caption{Histogram of MoM for the 8,000 most visited websites. The values were clipped at 200\% to improve plot readability.}
  \Description{The distribution is heavy-tailed, with a peak slightly above 0. It ranges from -100\% to 200\%.}
\label{fig:mom-hist}
\end{figure}

SimilarWeb provides information on how much the monthly visits of a domain increased or decreased in the last month compared to the previous one. This change, represented as a percentage, is called month-over-month variation (MoM). Due to practical data access issues, we could only analyze the MoM distribution of the 8,000 most visited domains. Fig. \ref{fig:mom-hist} shows that, despite a moderate variance (68\% of the MoMs stay between -5\% and 19\%), outliers are frequent both at low and high MoMs, strongly affecting non-robust statistical measures like the mean and the standard deviation. The average absolute difference between consecutive monthly visits is 15\%, about three times smaller than the average difference between SimilarWeb and NetMonitor measurements \cite{Prantl2018}. This slight variation indicates that methodological differences in traffic measurement are more important sources of error than differences in the time frame.

We also see that the distribution favors positive MoMs -- the distribution median is 3,6\%, and bootstrapping shows this value is significantly greater than zero. We suggest three explaining factors for this increase in monthly visits over time:

\begin{enumerate}
\item the total traffic on the Web is increasing, with people spending more time or more people gaining access to it \cite{ITU2023};
\item the top ranking websites are concentrating traffic, at the expense of less visited ones \cite{IntSoc2019}; and
\item there is a seasonal pattern (similar to the ones in \cite{Liu2017, Liu2018}) in which traffic typically increases from September to October.
\end{enumerate}
Further quantitative exploration of these factors would require data distributed over a larger time frame.

\begin{figure}[t]
\centering
\includegraphics[width=\columnwidth]{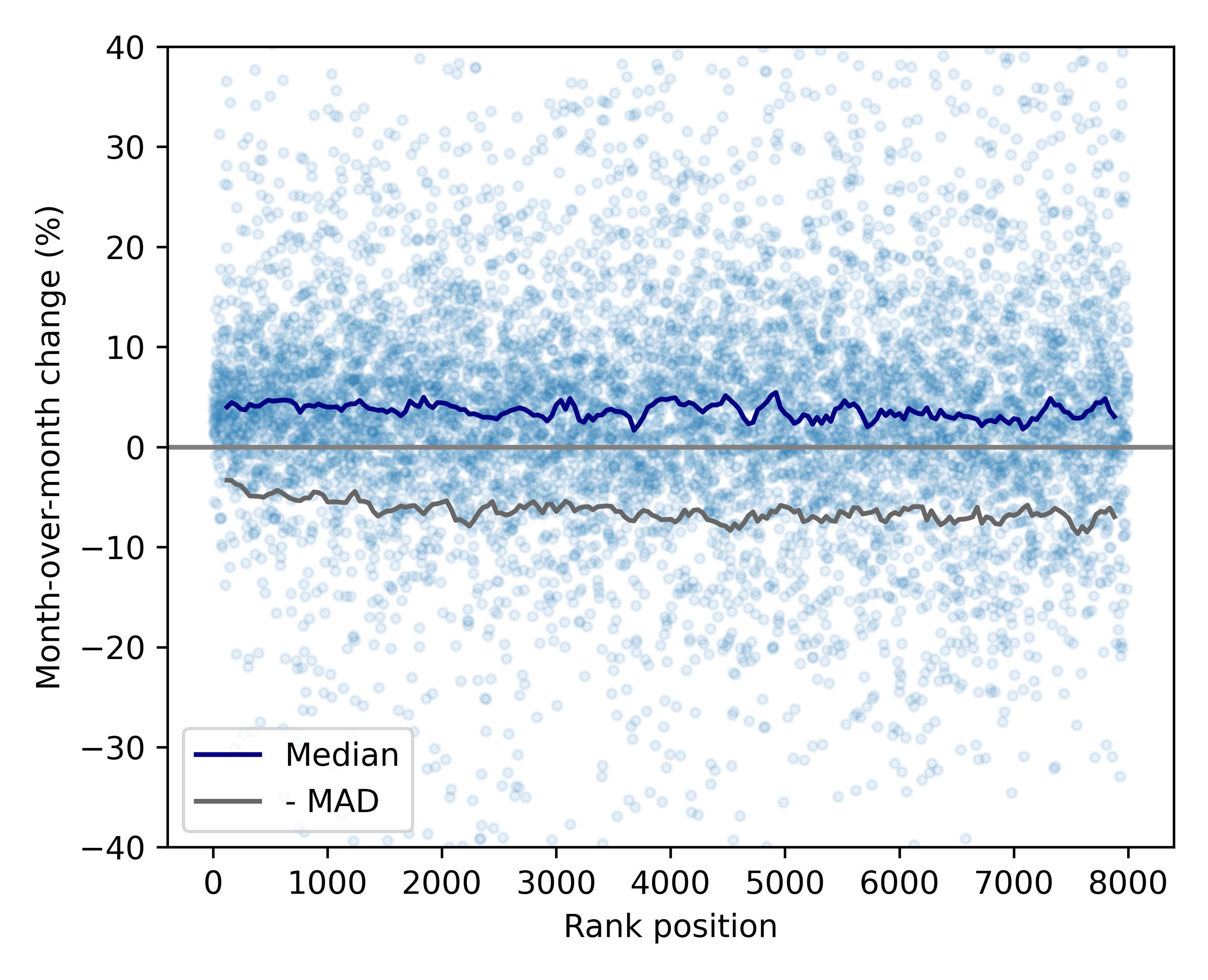}
\caption{MoM vs. rank position. The dots represent individual domains, and lines represent moving statistical measures. Outliers beyond $\pm 40\%$ are not shown.}
\Description{The dots are homogeneously scattered in the horizontal direction (rank position), with a vertical concentration around a MoM of 3.6\%. The lines representing the median and MAD are noisy but follow a mostly constant trend, except for the MAD at the top of the rank.}
\label{fig:mom-trend}
\end{figure}

Fig. \ref{fig:mom-trend} shows the MoM distribution as a function of rank position. The dark blue line represents the median computed in a moving window containing 200 domains, and the gray line represents the median absolute deviation (MAD) -- multiplied by $-1$ to avoid overlapping with the blue line -- computed in the same window. Apart from a statistically significant decrease in the MAD at the top of the rank (meaning that the monthly visits variation is more predictable for the top 2,000 websites), the distribution is reasonably stable in the remaining range.

Since the MoM measures fluctuations in a domain's visits and its distribution is stable for most of the rank, we adopted a centered (i.e., median-subtracted) version of the MoM distribution shown in Fig. \ref{fig:mom-hist} as a model for the statistical error on the monthly visits estimates, assuming the errors are independent and identically distributed. We will use this error model in the following section.

\subsection{Traffic distribution}
\label{sec:traffic}

We identified that the traffic (i.e., monthly visits) is distributed among domains in a power-law-like fashion. Previous papers have reported similar findings regarding the number of connections between internet nodes \cite{Faloutsos1999}, number of incoming links to a website \cite{Adamic2000}, the time spent by users on websites \cite{Crichton2021}, and number of unique visitors to domains \cite{Adamic1999, Webster2002}. Despite being correlated, these quantities are not the same. Among these studies, \cite{Adamic1999} had the most similar metric and data to ours: it used network-centric data from 60,000 AOL users active in December 1997 to rank 120,000 domains concerning the number of unique visitors.

\begin{figure}[t]
  \centering
  \includegraphics[width=\columnwidth]{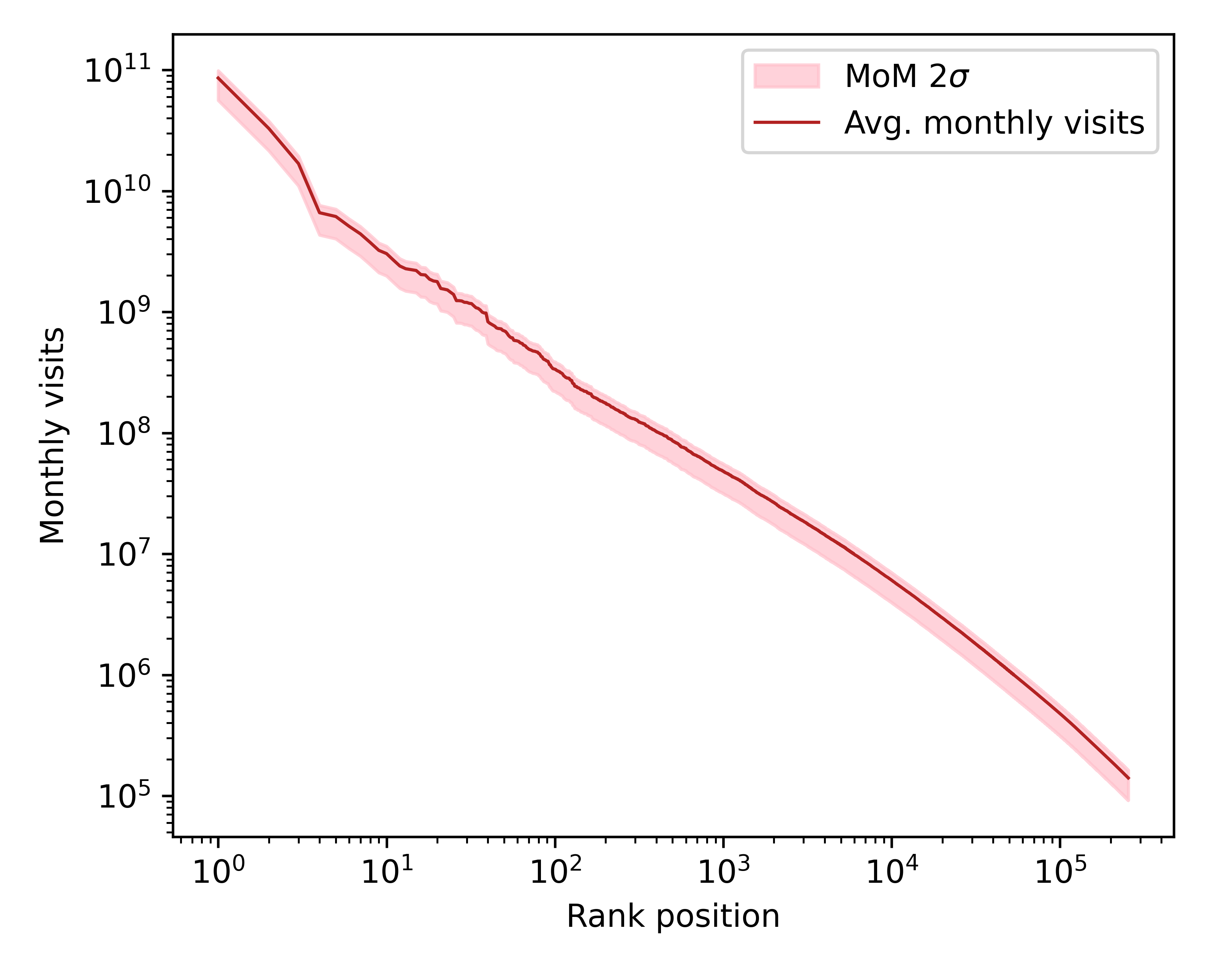}
  \caption{Average monthly visits as a function of the domain's position in the rank. Light bands represent a $2\sigma$ variation in the visits from month to month.}
  \Description{The plot employs log-log scales, and the average monthly visits vs. rank position forms an almost straight line. The band representing month-to-month variations is thin compared to its extension.}
\label{fig:traffic-dist}
\end{figure}

Fig. \ref{fig:traffic-dist} shows a log-log plot of the monthly visits $V$ vs. rank position $p$ for the domains in our dataset. A power-law $V(p)=V_0p^{\beta}$ appears as a straight line in such a plot. We obtain a best-fit model to our data with $V_0=4.1 \times 10^{11}$ and $\beta=-1.19$. We also represent as pink bands the range covered by a $2\sigma$-equivalent interval (i.e., from the 2nd to the 98th percentile) assuming statistical fluctuations described in Sec. \ref{sec:mom}. The narrowness of the bands demonstrates that the properties of the traffic distribution should be reasonably stable over time (at least for a few months) and that the overall ranking of the domains is not significantly affected by measurement errors.

\begin{figure}[t]
  \centering
  \includegraphics[width=\columnwidth]{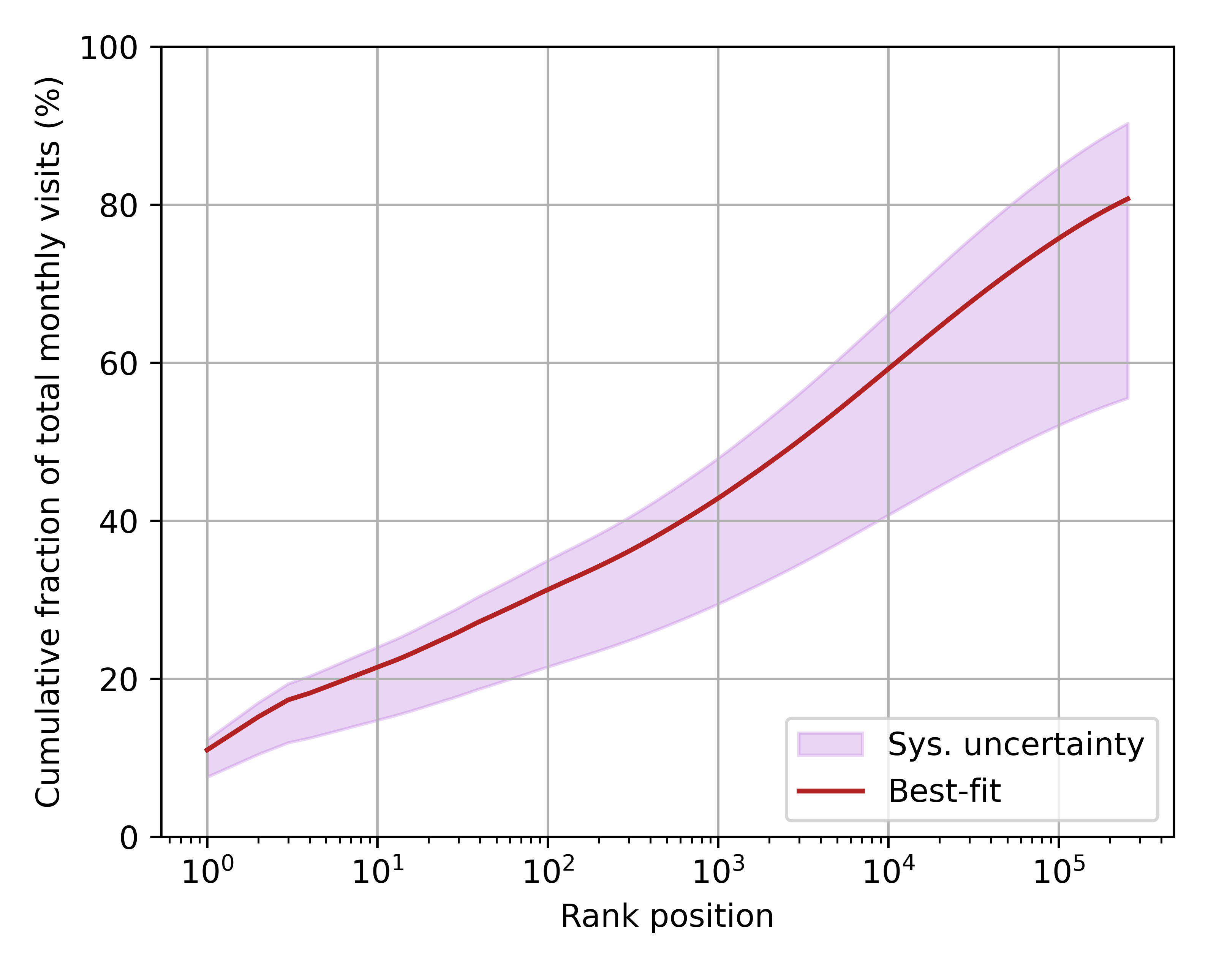}
  \caption{Best estimate of the cumulative traffic share of domains as a function of position in the rank (red line). The violet band represents the systematic uncertainty.}
  \Description{The best-fit trend grows continuously and passes 50\% at position  2,938, reaching 81\% at the last position. The width of the systematic uncertainty band grows with position. At the right end, it goes from 57\% of the total traffic share to 90\%.}
\label{fig:traffic-share}
\end{figure}

To calculate the share of the total web traffic captured by each domain in our dataset, we first need an estimate of this total traffic. For that, we extrapolated the best-fit power law to position 354 million, the number of registered domains as of June 2023 \cite{Verisign2023}, and aggregated the measured traffic along with the estimated traffic beyond position 254,661, resulting in a total of 781 billion visits per month. With 5.35 billion internet users as of January 2024 \cite{Petrosyan2024}, this amounts to 4.9 domain visits per person per day, a reasonable ballpark figure: previous work has shown that users from a metropolitan area in the United States visited an average of 20.1 websites per day on their computers, while the least active user (out of 257) visited an average of 2.9 \cite{Crichton2021}. Including rural regions and other countries should lead to a lower average. Fig. \ref{fig:traffic-share} shows the estimated traffic share for the 254,661 domains in our data.

Extrapolations are risky as there is no guarantee that the data behavior will remain the same beyond the measured range. However, it seems reasonable to assume that the monthly visits will continue to decline more or less like a power law, with an exponent $\beta$ not much different from the observed one. With this assumption, we expect the traffic contribution from the less visited domains to be minor, especially if $\beta < -1$.

To estimate a systematic uncertainty on the shares of total monthly web traffic, we fitted a power law up to position $p_{i}=10(p_{\mathrm{last}}/10)^{i/99}$, with $i=0,\ldots,99$ and $p_{\mathrm{last}}=254,661$, representing several hypothetical cases where we have data only for the first $p_i$ domains. The smallest and largest $\beta$ obtained, along with the respective power-law scaling factor $V_0=V(p_{\mathrm{last}}) / p_{\mathrm{last}}^{\beta}$, were used as alternative extrapolations to estimate a systematic uncertainty interval on the total monthly web traffic. The resulting interval for the traffic shares is shown as a violet band in Fig. \ref{fig:traffic-share}.

% Aqui mudei o prompt do ChatGPT.

Our analysis reveals a significant concentration of web traffic in a few domains. Our best estimate suggests that 50\% of all web traffic is directed to the top 3,000 domains, and 80\% is allocated to domains in our dataset. Even our most conservative estimate indicates considerable concentration, with 50\% of the traffic directed to approximately 60,000 domains. This scenario corresponds to a lower bound on the Gini coefficient -- a measure of inequality commonly used in economic analysis -- of 82\%, a value comparable to wealth inequalities observed in countries such as El Salvador, India, Kenya, Paraguay, and Turkey \cite{CreditSuisse2022}.

Table \ref{tab:share-stats} presents values for our best traffic extrapolation and for the systematic uncertainty boundaries with the smallest and largest traffic concentration, denoted as Sys. $-$ and Sys. $+$. These values are: the power-law exponent, $\beta$; the total monthly web traffic, $\sum V$; the average number of domain visits per person per day, $\bar{u}$; the extrapolated traffic for the last domain in the rank (among 354 million), $V_{\mathrm{min}}$; the number of top domains that accumulate 50\% of total web traffic, $p_{50\%}$; and the Gini inequality coefficient.

\begin{table}[ht]
  \caption{Main monthly traffic share attributes under different traffic extrapolations.}
  \label{tab:share-stats}
  \begin{tabular}{lcccccc}
    \midrule
    Extrap.   & $\beta$  & $\sum V$           & $\bar{u}$ & $V_{\mathrm{min}}$ & $p_{50\%}$ & Gini\\
    \midrule
    Sys. $-$  & -0.83    & $1.14\cdot 10^{12}$ &  7.1      & 341.1           & 62,621    & 82\% \\
    Best      & -1.19    & $7.81\cdot 10^{11}$ &  4.9      & 27.3            & 2,938     & 97\% \\
    Sys. $+$  & -1.52    & $6.98\cdot 10^{11}$ &  4.3      & 2.4             & 1,342     & 99\% \\
    \midrule
  \end{tabular}
\end{table}

The $V_{\mathrm{min}}$ value for the Sys. $-$ boundary is implausibly large, indicating that the traffic share must decline faster than this rate, especially towards the end of the traffic rank. Since not all 354 million domains have an associated website, traffic must decline even faster to reach a reasonable $V_{\mathrm{min}}$ at an earlier position. Thus, the actual traffic share is likely closer to our best estimate than to its lower boundary. The Gini coefficient associated with our best traffic extrapolation far exceeds the wealth inequality of any country worldwide. The fact that the domains in our dataset likely represent the destination of 80\% of all web traffic makes them an excellent sample of the current state of the Web, at least in terms of content access rate.

\subsection{Popular web industries}
\label{sec:content}

SimilarWeb categorizes domains into 210 different industry segments. Some of these segments are broader categories that encompass others. By considering only the finer, non-overlapping categories, we are left with 187 industry segments. We aggregated the monthly visits of domains into these 187 categories to estimate the most popular industries on the Web. This computation better represents the popularity of topics than others found in the literature (e.g., \cite{Agarwal2022}), as it ranks them by monthly visits rather than by the number of websites. Additionally, it utilizes a significantly vaster set of domains compared to previous studies. 

\begin{figure}[t]
  \centering
  \includegraphics[width=\columnwidth]{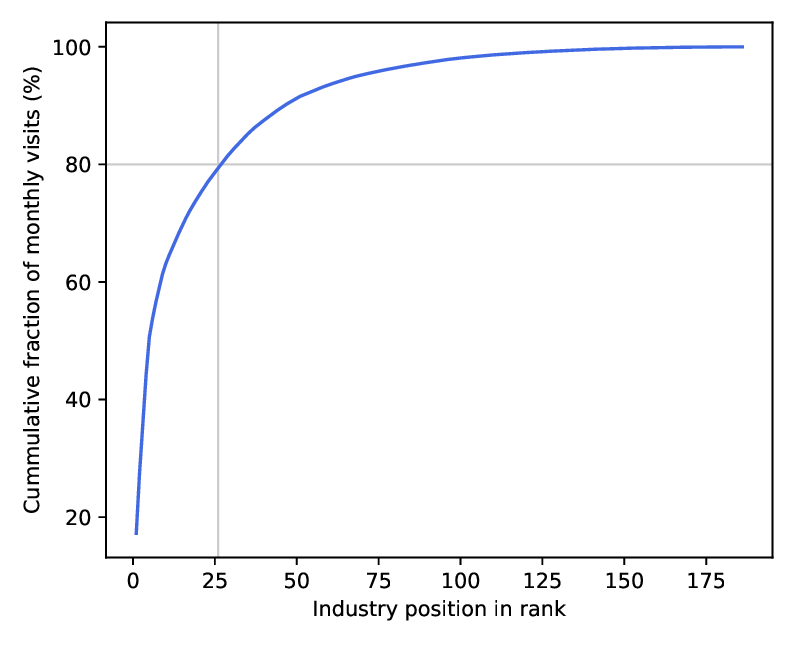}
  \caption{Cummulative share of monthly visits aggregated by web industry, from the most to the least popular.}
  \Description{The cumulative share curve is very smooth, growing rapidly at the start and flattening out when two-thirds of the industries are accounted for.  There is an ``elbow'' at industry position 26, where it reaches the cummulative share of 80\%.}
\label{fig:cum-industry}
\end{figure}

\begin{figure}[t]
  \centering
  \includegraphics[width=\columnwidth]{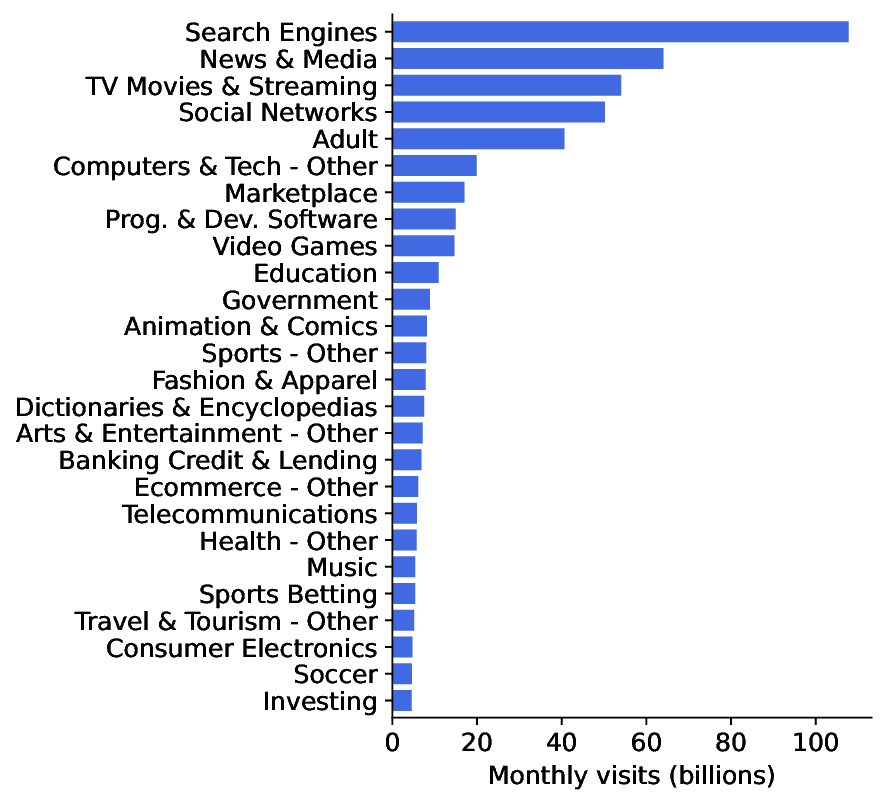}
  \caption{Monthly visits to the top 26 industries. We shortened some industry names to save space on the plot.}
  \Description{Bar plot with Search Engines leading the visits. With 1/3 fewer visits, News and Media appear, followed by TV, movies, and Streaming, then by Social Networks and Adults. The remaining categories start with half the Adult visits, and the visits distribution almost flattens out.}
\label{fig:industries-visits}
\end{figure}

Figure \ref{fig:cum-industry} illustrates that traffic coalesces in a few industries, with 50\% directed to just five segments and 80\% directed to 26 segments. Figure \ref{fig:industries-visits} depicts the average monthly visits for these sectors. In most cases, despite the power-law distribution of traffic within industries, monthly visits are not monopolized by a single domain or company, and the most popular domains accumulate at most 15\% of their industry's traffic. However, there are exceptions, as noted below.  

Figure \ref{fig:industries-visits} demonstrates the prevalence of ``Search Engines'' as the most visited industry. This prevalence is expected due to the lack of a built-in index on the Web, making search engines entry points for most users. This trend is further amplified by the integration of address and search bars in web browsers, leading users to search even for known websites instead of directly typing their URLs \cite{Cannon2008}. In this industry, google.com dominates with 79\% of the traffic (considering the top 10,000 domains in the segment). The following competitors are the Chinese search engine baidu.com, with 4.7\%, and the Russian search engine yandex.ru, with 3.0\%.

With 40\% less traffic than ``Search Engines,'' the next most popular industry is ``News and Media,'' which comprises a wide variety of websites, including web portals such as yahoo.com. Following this is "TV Movies and Streaming," where youtube.com (owned by Alphabet, like google.com) captures 61\% of the industry's traffic. This segment is succeeded by ``Social Networks and Online Communities,'' another case where the industry's traffic predominantly flows to a single company. When combining facebook.com, instagram.com, and whatsapp.com, Meta receives 52\% of the segment's traffic. The last industry in the top five, before a traffic drop of more than 50\%, is ``Adult.'' Despite its popularity on the Web, this segment may be underrepresented in surveys and research where subjects know they are being monitored \cite{NIC2022, Crichton2021}. Among the remaining 26 industries listed in Figure \ref{fig:industries-visits}, only ``Dictionaries and Encyclopedias'' has a domain (wikipedia.org) with more than half (56\%) of the total industry's traffic.

It is essential to observe that domains categorized as ``Search Engines,'' such as google.com, yandex.ru, and baidu.com, offer more services than just web search. Consequently, a portion of the traffic attributed to the ``Search Engines'' segment pertains to other industries listed in SimilarWeb's classification, such as ``Email,'' ``File Sharing and Hosting,'' and ``Maps.'' While some of these services are hosted on subdomains and are tracked independently by SimilarWeb (e.g., news.google.com and cloud.google.com), others are not (e.g., support.google.com, mail.google.com, and calendar.google.com). Nevertheless, it is feasible to disregard traffic from all subdomains \cite{SimilarWeb2024b}. For google.com, this results in a 15\% reduction in traffic, maintaining the integrity of our rankings. Note that services like Google Maps are hosted at www.google.com/maps, so their data remains inseparable from that of google.com.

\begin{figure}[t]
  \centering
  \includegraphics[width=\columnwidth]{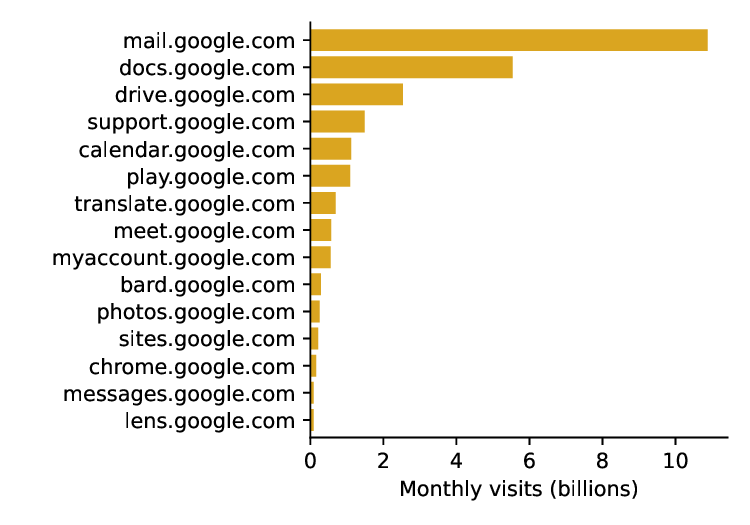}
  \caption{Traffic on Google's subdomains not tracked independently by SimilarWeb and unrelated to search.}
  \Description{Horizontal bar plot shows the average monthly visits to Google's subdomains. mail.google.com has the most visits: 10.9 billion. It is followed by docs.google.com (5.5 billion), drive.google.com (2.5 billion), support.google.com (1.5 billion), calendar.google.com (1.1 billion), play.google.com (1.1 billion), translate.google.com (694 million), and meet.google.com (570 million). There are seven more subdomains in the plot.}
\label{fig:google-visits}
\end{figure}

Considering Google's widespread popularity and the multitude of services it offers, it is valuable to assess the significance of each service within the overall traffic of google.com and its effect on the traffic attributed to the respective industries. To achieve this, we conducted manual searches on SimilarWeb to gather data on Google's subdomains that are neither independently tracked nor related to search. Fig. \ref{fig:google-visits} illustrates these findings. By juxtaposing these subdomains' traffic data with google.com's average monthly visits of 85.7 billion, it becomes evident that search continues to reign as Google's most frequented service on the Web.

Nonetheless, Google offers several additional services that would significantly contribute to its industry's traffic if SimilarWeb had independently monitored them. For instance, in the ``Email'' segment, mail.google.com received an average of 10.9 billion monthly visits, whereas the top domain in this segment, live.com, received only 2 billion visits. Similarly, with 5.6 billion visits, docs.google.com would lead the ``Programming and Developer Software'' segment, which includes Software as a Service (SaaS) platforms like office.com, which received 1.5 billion visits. Additionally, drive.google.com would lead the ``File Sharing and Hosting'' segment with 2.5 billion visits, surpassing the 186 million visits received by the industry leader, mediafire.com. In the ``Computers Electronics and Technology - Other'' category, domains like calendar.google.com, meet.google.com, and bard.google.com would rank among the top ten most visited domains. Furthermore, translate.google.com would secure the third position in the ``Dictionaries and Encyclopedias'' segment, surpassing the translation website deepl.com.

\subsection{Manual inspection of top domains}
\label{sec:top-domains}

SimilarWeb data provides no other information about the domains besides visitation metrics and the domain's industry. To address this gap, we manually inspected and researched the 116 most visited domains, which collectively capture between 22\% and 36\% of the total web traffic (with a best estimate of 32\%). Our objective was to answer the following questions about them:

\begin{enumerate}
\item Does the website primarily offer SaaS?
\item Does the website produce its content? For marketplaces, does the domain owner sell its own products on the site?
\item Does the website function as a platform for user-generated content? For e-commerce platforms, does the site allow third-party sellers to list their products? Comments on original content (such as those found on some news websites) were insufficient to classify the site as a platform. We did not consider news aggregators or similar sites that curate content from the Web as platforms since the content is not user-provided.
\item Does the website require users to log in before accessing its main content?
\item Does the website charge users for access to its main content and features? We ignored charges for additional features and, for marketplaces, the charges imposed on sellers.
\item Is the domain owner's primary business or activity related to the Web, or is a significant portion of their business conducted offline? We considered news websites primarily web-related, but not e-commerce sites that sell their own products.
\item Who is the ultimate owner of the domain? In the case of subsidiaries, we identified the ultimate parent company as the domain's final owner. For partial ownership, we only considered parent companies that owned at least 50\% of the subsidiaries.
\item Is the ultimate owner of the domain a for-profit organization?
\item In which country is the domain's ultimate owner located?
\end{enumerate}
We present our annotations in Tables \ref{tab:annot-domains-1} and \ref{tab:annot-domains-2} (see Appendix \ref{sec:annotations}), and in our repository.\footnote{\url{https://github.com/cewebbr/web-unpacked/blob/main/data/cleaned/domains-annotated\_v03.csv}} While the analysis of this subset of domains does not guarantee a comprehensive representation of total Web traffic, the trends observed within it are likely to extend to the following several thousand domains, thus reflecting the characteristics of a significant portion of Web usage.

Among the top 116 listed domains, Microsoft is the largest owner with 11 domains. These include linkedin.com, openai.com\footnote{While we attribute ownership of OpenAI to Microsoft, the precise nature of their relationship remains a topic of debate.}, bing.com, github.com, and several others associated with Microsoft itself and its SaaS products like office365.com. Following is Amazon, the owner of twitch.tv, imdb.com, and five Amazon-related domains. Alphabet holds ownership of five domains, including youtube.com and various Google domains, while Meta possesses four (facebook.com, instagram.com, whatsapp.com, and messenger.com). The first non-US company in terms of top domains is the Russian VK, which owns dzen.ru, vk.com, mail.ru, and ok.ru, followed by the Japanese SoftBank Group Corp., which owns three domains.

\begin{figure}[t]
  \centering
  \includegraphics[width=\columnwidth]{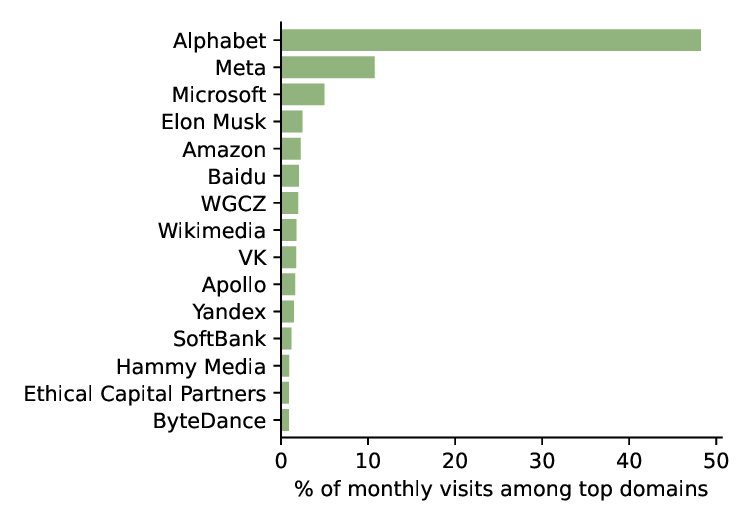}
  \caption{Fraction of the top 116 domains' total traffic aggregated by final owner. The plot shows only the top 15 final owners.}
  \Description{Horizontal bar plot. Alphabet is by far the owner with the most visits, reaching 49\% of the top domains' total traffic. Meta comes in second, with 11\%. Then we have Microsoft at 5\%, Elon Musk at 2.5\%, Amazon at 2.3\%, and Baidu at 2.1\%.}
\label{fig:owners-traffics}
\end{figure}

Suppose we rank the final owners by their aggregated traffic. In that case, Alphabet receives 120 billion visits monthly, accounting for almost 50\% of the total traffic of the top 116 domains (see Fig. \ref{fig:owners-traffics}). This ranking also roughly follows a power-law distribution, but its best-fit exponent, $\beta$, is more negative than the best-fit for the disaggregated top 116 domains: -1.22 versus -0.98. It is important to note that this decrease in $\beta$ is not an artifact caused by the aggregation itself. Randomly aggregating domains would typically blend famous and less known domains, evening out the traffic and resulting in a less negative $\beta$. Thus, there must be a socio-economic explanation for why the same companies own highly visited domains. Although advancing such an explanation is beyond the scope of this paper, we conjecture that this phenomenon partially stems from large companies acquiring popular websites (e.g., youtube.com, linkedin.com, dzen.ru, twitch.tv, imdb.com, and github.com were independent websites later purchased by their current owners).

When we aggregate the top domains' visits by the ultimate owner's country, the concentration becomes even more pronounced. The best-fit power-law exponent reaches $\beta=-2.00$, with the United States capturing 80\% of the total traffic of the top 116 domains. This dominance is followed by China (4.5\%), Russia (3.7\%), Japan (2.6\%), the Czech Republic (2.1\%), and South Korea (0.97\%).

\begin{figure}[t]
  \centering
  \includegraphics[width=\columnwidth]{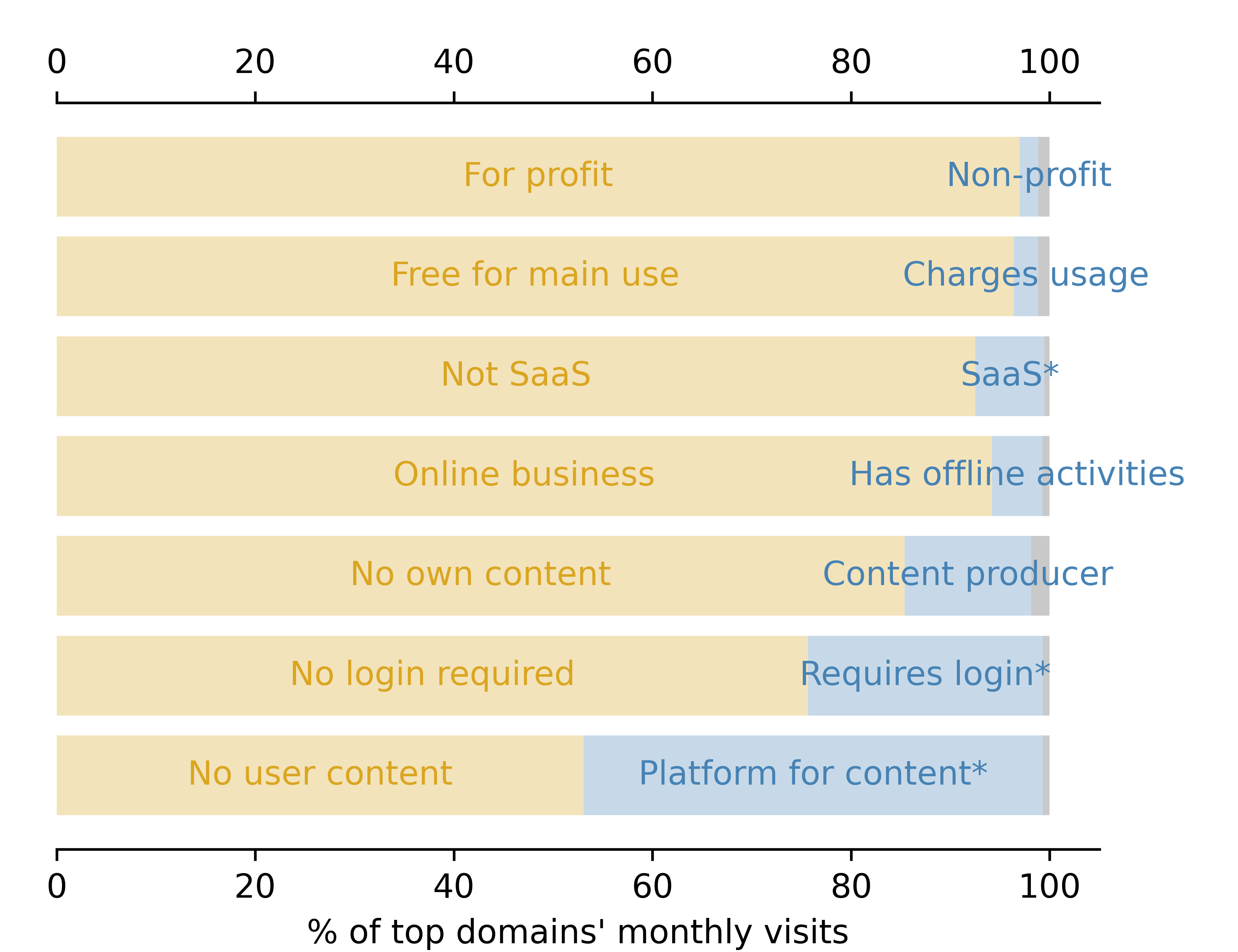}
   \caption{Fraction of the top 116 domains' total traffic directed to domains classified under seven binary properties. The gray segments denote traffic directed to unclassified domains. Segments marked with * would be larger depending on how we deal with large search engines (see text).}
  \Description{Each bar shows the traffic share directed to the domains' classes in a binary classification. The plot shows that more than 90\% of the top 116 domains' traffic is directed to domains owned by for-profit companies, that are free to access, that are nos SaaS, and that are mainly related to an online business (with no ``offline'' counterpart). We also see that about 85\% of the traffic share goes to domains that do not produce their own content, and almost 80\% do not require a login. About 47\% are platforms where users post their content.}
\label{fig:type-share}
\end{figure}

Fig. \ref{fig:type-share} summarizes the traffic shares associated with the binary answers to the remaining questions (1 to 6 and 8). Each bar in the plot illustrates how the traffic is distributed among domains subjected to a specific binary classification. Notably, the vast majority (97\%) of the traffic to the top 116 domains flows to domains owned by profit-seeking companies. Interestingly, among these 116 domains, only two are not-for-profit: wikipedia.org and archiveofourown.org. This observation is intriguing, especially considering that 96\% of the top traffic goes to websites that do not charge for access or their main functionalities. This seeming contradiction underscores emerging business models that rely on charging for additional features (e.g., the Freemium model \cite{Georgieva2015}), growing the user base for a later sell-out  \cite{Eocman2006}, and operating two-sided markets where users are subsidized to facilitate data collection, advertising, and sales for the other market side \cite{Hermes2020}. Additionally, the figure reveals that 94\% of the visits stream to web-related domains, indicating businesses with no offline counterpart (e.g., search engines, online advertising, marketplaces, social media, SaaS).

Figure \ref{fig:type-share} emphasizes that 85\% of the total traffic to top domains flows to platforms that do not generate their content, such as social media platforms, marketplaces, news aggregators, and search engines. Conversely, 46\% of the traffic goes to websites where users themselves create content. Excluding traffic to search engines, which neither produce content nor allow users to post content, the proportion of traffic allocated to user-generated content platforms rises to 77\%, with minimal impact on the distribution across other domain categorizations. This large percentage underscores the central role of user-content platforms in the Web's contemporary landscape.

Finally, the analysis reveals that SaaS does not emerge as a prominent traffic driver for top domains. This fact suggests that Web visitors prioritize content over services, indicating a lesser emphasis on the Web as a software platform. However, as detailed in Section \ref{sec:content}, notable domains like google.com, yandex.ru, baidu.com, amazon.com, and yahoo.com offer additional services beyond their primary functions, some of which fall under the SaaS category. For instance, approximately 15\% of Google's traffic is directed to subdomains primarily recognized as SaaS (refer to Fig. \ref{fig:google-visits}). A similar scenario unfolds regarding the necessity for user authentication to access the Web: while 76\% of the traffic to top domains goes to websites categorized as not requiring login credentials, a considerable portion of this traffic flows to domains with subdomains demanding authentication. Thus, at least 24\% of this traffic necessitates user authentication.

We employed clustering techniques to categorize the top 116 domains into four groups. We aimed to minimize the average Hamming distances within each group across a 7-dimensional binary feature space constructed from our questionnaire responses. The domains within each cluster exhibit similar characteristics, allowing us to delineate the clusters using the following archetypes:
\begin{enumerate}
\item \emph{SaaS}: This group comprises domains primarily characterized by SaaS offerings, devoid of offline activities and requiring user authentication. While some domains charge users for their core services, many do not. Examples include openai.com, zoom.us, office365.com, and canva.com. Our clustering process assigned 12 domains to this group.
\item \emph{Open content providers}: Domains in this cluster serve as content producers, offering access to their content without imposing a paywall or requiring user authentication. Notably, this is the only group to include businesses with offline operations and closely resembles the ethos of the early internet era, often referred to as ``Web 1.0.'' Examples encompass samsung.com, walmart.com, bbc.co.uk, weather.com, and yahoo.com. Amazon.com was categorized within this group by our clustering algorithm, along with other 31 domains.
\item \emph{Platforms}: This cluster encompasses domains whose activities are confined to the online realm, where users contribute the entirety of the content, and access to the websites is free. Prominent examples comprise youtube.com, roblox.com, chaturbate.com, tiktok.com, wikipedia.org, and booking.com. Interestingly, search engines and messaging web services such as telegram.org and discord.com were assigned to this group by our clustering method. In total, 67 domains were classified under this archetype.
\item \emph{Subscription content providers}: Domains in this category are online content producers that mandate user authentication and levy charges for accessing their content. Examples include nytimes.com, espn.com, netflix.com, and disneyplus.com. Our clustering process identified five domains within this group.
\end{enumerate}

\begin{figure}[t]
  \centering
  \includegraphics[width=\columnwidth]{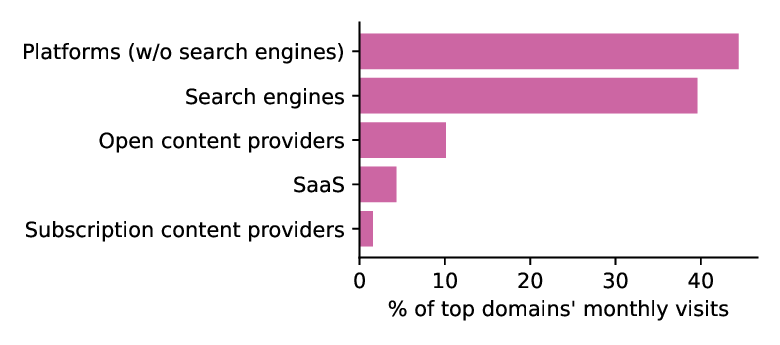}
  \caption{Fraction of the top 116 domains' total traffic directed to each of the 4 clusters. Search engines were manually removed from cluster group ``Platforms''.}
  \Description{Horizontal bar plot. 44\% of the top 116 domains' traffic goes to Platforms. This is followed by search engines (40\%), Open content providers (10\%), SaaS (4.3\%), and Subscription content providers (1.6\%).}
\label{fig:cluster-share}
\end{figure}

Figure \ref{fig:cluster-share} shows that the cluster designated as ``Platforms'' is the primary destination for traffic among the top 116 domains, even when excluding search engines from consideration. While some traffic directed toward search engines could be labeled as SaaS due to their supplementary services, the preeminence of the Platforms cluster remains unassailable.

\section{Summary and conclusions}
\label{sec:conclusions}

%Our study relied on data from SimilarWeb, a company that estimates web traffic for competitive analysis purposes. SimilarWeb combines user-centric, server-centric, and network-centric measurements to provide comprehensive insights into global web usage \cite{SimilarWeb2024c}. Access to representative user-centric data is crucial for generating an impartial depiction of the most visited websites, and SimilarWeb stands out as a leading source for such data \cite{Diachuk2021, Hardwick2018, Jansen2022, Prantl2018}.

Our analysis commenced with examining the month-to-month fluctuations in websites' monthly visit counts to assess the stability of website traffic rankings (Section \ref{sec:mom}). We observed that these fluctuations can be substantial, mainly due to the presence of outliers (see Fig. \ref{fig:mom-hist}). However, the ranking remains relatively stable in the short term, largely owing to its power-law-like distribution segregating websites by orders of magnitude (refer to Fig. \ref{fig:traffic-dist}). Notably, our traffic evaluation represents, to the best of our knowledge, the most comprehensive assessment published to date, encompassing data from 254,661 domains.

By utilizing a power-law fit to extrapolate monthly visits from rank position 254,661 to encompass the estimated 354 million existing domains as of 2023 \cite{Verisign2023}, we projected total web traffic to reach 781 billion visits per month (Section \ref{sec:traffic}). This projection corresponds to an average of 4.9 visits per user per day, a figure broadly in line with existing estimates \cite{Crichton2021, Petrosyan2024}. Such an extrapolation enabled us to gauge the degree of concentration of web usage within a small subset of top websites. Our analysis suggests that approximately 50\% of all web traffic is directed towards roughly 3,000 websites, with 80\% of the total traffic accounted for by the websites in our dataset (refer to Fig. \ref{fig:traffic-share}). This degree of concentration is remarkably high, as indicated by a Gini coefficient of 97\% (see Table \ref{tab:share-stats}).

To the best of our knowledge, our estimation of web usage concentration is the only published one to date. Its implications for web studies are profound, as it underscores, through quantitative data, that the characteristics of the Web, particularly in terms of usage, can be reasonably inferred from a minimal fraction of all registered domains. While our dataset covers a mere 0.07\% of all domains, it is likely to encapsulate around 80\% of total web traffic.

Using this dataset, we revealed an accentuated concentration of user interest over a limited number of industries (refer to Fig. \ref{fig:cum-industry}). Specifically, we found that half of the traffic in our dataset flows toward five industries: ``Search Engines,'' ``News and Media,'' ``TV Movies and Streaming,'' ``Social Networks and Online Communities,'' and ``Adult'' (see Fig. \ref{fig:industries-visits}). Previous studies examining user interests have often relied on smaller, biased samples \cite{Crichton2021}. It is worth noting that certain top domains, such as google.com, yandex.ru, baidu.com, amazon.com, and yahoo.com, encompass multiple services that could be classified under different industries than their primary ones. For instance, an analysis of Google's traffic reveals that approximately 15\% of it is directed to subdomains, with a significant portion associated with SaaS, particularly in functionalities like email, document editing, and file storage (see Fig. \ref{fig:google-visits}). Consequently, the actual usage attributed to search engines may be slightly lower, while the utilization of SaaS-related industries may be slightly higher than depicted in Fig. \ref{fig:industries-visits}.    

Our in-depth analysis of the 116 most visited domains, collectively responsible for approximately one-third of all web traffic, reveals that nearly all of this traffic goes towards websites owned by for-profit digital tech companies. This observation highlights the Web primarily as a vehicle for commercial enterprise. Notably, this enterprise predominantly manifests as purely online ventures rather than online extensions of offline activities. Originally conceived as an open and uncharted domain, the Web has evolved into a realm ripe for commercial exploitation.

Furthermore, it is noteworthy that almost all of the 116 domains analyzed do not charge users for access or the core functionalities offered by them. This lack of paywalls highlights prevalent business models on the Web, which rely on alternative revenue streams. Such revenue generation methods may include charging for supplementary features, as observed in domains like google.com, youtube.com, roblox.com, amazon.com, and github.com. Additionally, some companies secure venture capital funding to pursue a ``growth at all costs'' strategy, as exemplified by platforms such as whatsapp.com and quora.com \cite{Kutcher2014, ErnstYoung2014, Constine2013}. Moreover, certain companies engage in two-sided markets, where the user base serves more as an asset than traditional customers, providing valuable data, customers, and audiences for the other side of the market. Prominent examples of this approach include facebook.com, booking.com, ebay.com, and youtube.com \cite{Zuboff2019}.

Most traffic to top domains goes towards websites that do not generate their content. Instead, this content originates from various sources, including users (e.g., instagram.com, xvideos.com, github.com, and messenger.com), sellers (ebay.com, booking.com, and aliexpress.com), or other external websites (such as search engines and news aggregators). We refer to websites that exhibit these predominant characteristics as ``Platforms.'' Platforms compose most of the top domains (67 out of 116) and collectively capture over 80\% of its traffic (refer to Fig. \ref{fig:cluster-share}), underscoring their significance in scholarly discourse \cite{Hein2020, Hermes2020, Zuboff2019}.

While our platform identification efforts focused on approximately one-third of the web traffic, our data suggests that platform hegemony could extend to more significant portions of web usage. Several prominent industries outlined in Fig. \ref{fig:industries-visits} primarily consist of such websites. For instance, ``Search Engines,'' ``Social Networks and Online Communities,'' and ``Adult'' industries are mostly comprised of platforms (refer to Tables \ref{tab:annot-domains-1} and \ref{tab:annot-domains-2}). Additionally, ``TV Movies and Streaming,'' although primarily constituted by subscription content providers, sees its traffic largely dominated by youtube.com, a platform. While platform dominance has been previously acknowledged in the financial realm \cite{Evans2016} and may be readily observed through personal experience, we are unaware of former quantitative evidence of its prevalence over web traffic.

The concentration of web visits on platforms can be seen as a form of privatization of the Web, particularly concerning its usage dynamics. Despite the abundance of domain names and websites -- publishing an independent website remains straightforward and accessible to many -- visits are a scarce resource that platforms have enclosed. Within the web context, visits represent more than mere audience numbers, as visitors actively contribute to the content hosted on platforms. Web usage, encompassing activities such as viewing, posting, and interacting, predominantly occurs within these platforms, which dictate the rules governing these activities and determine the fate of the content and data generated through these visits.

Our analysis of the top 116 domains also reveals that at least 24\% of their visits necessitate user authentication, indicating that while the Web appears largely open, a noticeable portion operates behind login barriers. Furthermore, our findings suggest that other uses than SaaS may predominate on the Web. However, rather than relying on a web survey like ours, a dedicated application survey would be necessary to ascertain whether applications are transitioning to the Web.

Additionally, we observe that a staggering 80\% of the traffic to the top domains flows to companies in the United States. This substantial concentration, which could potentially characterize the Web as predominantly an American enterprise, has been previously highlighted in terms of companies' market value \cite{Evans2016}, but not in terms of traffic distribution. As discussed in Section \ref{sec:top-domains}, the accumulation of traffic on specific countries necessitates a socio-economic and political explanation. Hermes \emph{et al.} took a step in this direction by interviewing European experts and top managers to glean insights into the reasons behind the dominance of American platforms in terms of market value \cite{Hermes2020b}. Their findings suggest that factors such as a results-oriented mindset, willingness to take risks, a sizable domestic market, state investment, early-mover advantage, the establishment of technology hubs (such as Silicon Valley), close collaborations with universities, and access to venture capital collectively contribute to American dominance in the web landscape.

The utilization of the Web exhibits, of course, significant diversity, varying among different individuals, social groups, and countries. However, our analysis indicates that, currently, on a global scale, the Web is predominantly characterized by the overwhelming presence of for-profit American platforms with business models not reliant on subscription fees. This observation provides quantitative support for common descriptions of the digital realm, such as ``Platform capitalism'' \cite{Srnicek2016}, ``Surveillance capitalism'' \cite{Zuboff2019}, and ``Technofeudalism'' \cite{Varoufakis2024}. While these descriptions often focus on corporate giants like Alphabet, Meta, Microsoft, and Amazon, potentially conveying that their findings are specific to these companies, our analysis suggests otherwise. These companies not only command a significant portion of web usage but also exhibit characteristics mirrored by numerous smaller entities, indicating that their descriptions may reflect broader trends across the web ecosystem.

% Padronizar Figure -> Fig.

\bibliographystyle{ACM-Reference-Format}
\bibliography{main}

\balance

\appendix

\section{Annotated data}
\label{sec:annotations}

Tables \ref{tab:annot-domains-1} and \ref{tab:annot-domains-2} provide data about the top 116 domains in terms of web traffic. The first three columns present information from SimilarWeb: the domain name, its corresponding industry (with some industry names abbreviated for table readability), and the average monthly visit count (measured in billions), denoted as $V$. Subsequent columns, except for the last one, were manually annotated based on the questions outlined in Section \ref{sec:top-domains}, with binary responses marked as 1 for "yes" and 0 for "no."

Tables \ref{tab:annot-domains-1} and \ref{tab:annot-domains-2} provide data about the top 116 domains in terms of web traffic. The first three columns present information from SimilarWeb: the domain name, its corresponding industry (with some industry names abbreviated for table readability), and the average monthly visit count (measured in billions), denoted as $V$. Subsequent columns, except for the last one, were manually annotated based on the questions outlined in Section \ref{sec:top-domains}, with binary responses marked as 1 for "yes" and 0 for "no." 

The first three specify if we considered the respective domain: a SaaS provider (question 1); a content producer, $CP$ (question 2); and a platform for user-generated content, $Pl$ (question 3). The following three columns inform if the domain: requires user login, $Log$ (question 4); charges for primary usage or access, $Ch$ (question 5); and primarily operates as a web-related business (question 6). Subsequent columns include information on the final owner of the domain according to Wikipedia (question 7, with some owner names abbreviated for table presentation), whether the owner is a for-profit organization, $\$$ (question 8), and the country of origin of the owner (question 9, with some country names abbreviated or transformed into acronyms for table presentation). The last column indicates the cluster assignment (out of four) determined through unsupervised agglomerative clustering. We defined the cluster descriptors after examining the prevailing characteristics of each cluster: "Plat" for platforms; "OCP" for open content providers; "SaaS" for software as a service; and "SCP" for subscription content providers.

\balance

\begin{table*}
\scalebox{0.88}{
  \begin{tabular}{llrllllllllll}
\toprule
             Domain &                      Industry &  V ($10^9$) & SaaS & CP & Pl & Log & Ch & Web &                    Owner & \$ &     Country &   Cl \\
\midrule
         google.com &                Search Engines &      85.730 &    0 &  0 &  0 &   0 &  0 &   1 &                 Alphabet &  1 &         USA & Plat \\
        youtube.com &        TV Movies \& Streaming &      32.810 &    0 &  0 &  1 &   0 &  0 &   1 &                 Alphabet &  1 &         USA & Plat \\
       facebook.com &               Social Networks &      16.810 &    0 &  0 &  1 &   1 &  0 &   1 &                     Meta &  1 &         USA & Plat \\
      instagram.com &               Social Networks &       6.602 &    0 &  0 &  1 &   1 &  0 &   1 &                     Meta &  1 &         USA & Plat \\
        twitter.com &               Social Networks &       6.143 &    0 &  0 &  1 &   1 &  0 &   1 &                Elon Musk &  1 &         USA & Plat \\
          baidu.com &                Search Engines &       5.095 &    0 &  0 &  0 &   0 &  0 &   1 &                    Baidu &  1 &       China & Plat \\
      wikipedia.org & Dictionaries \& Encyclopedias &       4.413 &    0 &  0 &  1 &   0 &  0 &   1 &                Wikimedia &  0 &         USA & Plat \\
          yahoo.com &                 News \& Media &       3.747 &    0 &  1 &  0 &   0 &  0 &   1 &                   Apollo &  1 &         USA &  OCP \\
          yandex.ru &                Search Engines &       3.225 &    0 &  0 &  0 &   0 &  0 &   1 &                   Yandex &  1 &      Russia & Plat \\
       whatsapp.com &               Social Networks &       3.033 &    1 &  0 &  1 &   1 &  0 &   1 &                     Meta &  1 &         USA & Plat \\
        xvideos.com &                         Adult &       2.676 &    0 &  0 &  1 &   0 &  0 &   1 &                     WGCZ &  1 &  Czech Rep. & Plat \\
         amazon.com &                   Marketplace &       2.394 &    0 &  1 &  1 &   0 &  0 &   0 &                   Amazon &  1 &         USA &  OCP \\
        pornhub.com &                         Adult &       2.273 &    0 &  0 &  1 &   0 &  0 &   1 & Ethical Capital Partners &  1 &      Canada & Plat \\
         tiktok.com &               Social Networks &       2.235 &    0 &  0 &  1 &   0 &  0 &   1 &                ByteDance &  1 &       China & Plat \\
           xnxx.com &                         Adult &       2.199 &    0 &  0 &  1 &   0 &  0 &   1 &                     WGCZ &  1 &  Czech Rep. & Plat \\
        yahoo.co.jp &                 News \& Media &       2.035 &    0 &  1 &  0 &   0 &  0 &   1 &                 SoftBank &  1 &       Japan &  OCP \\
           live.com &                         Email &       2.019 &    1 &  0 &  0 &   1 &  0 &   1 &                Microsoft &  1 &         USA & SaaS \\
         reddit.com &               Social Networks &       1.863 &    0 &  0 &  1 &   0 &  0 &   1 &                   Reddit &  1 &         USA & Plat \\
       docomo.ne.jp &            Telecommunications &       1.801 &    0 &  1 &  0 &   0 &  0 &   0 &                      NTT &  1 &       Japan &  OCP \\
       linkedin.com &               Social Networks &       1.780 &    0 &  0 &  1 &   1 &  0 &   1 &                Microsoft &  1 &         USA & Plat \\
         openai.com &     Computers \& Tech - Other &       1.561 &    1 &  1 &  0 &   1 &  0 &   1 &                Microsoft &  1 &         USA & SaaS \\
         office.com &        Prog. \& Dev. Software &       1.541 &    1 &  0 &  0 &   1 &  0 &   1 &                Microsoft &  1 &         USA & SaaS \\
       xhamster.com &                         Adult &       1.519 &    0 &  0 &  1 &   0 &  0 &   1 &              Hammy Media &  1 &      Cyprus & Plat \\
        netflix.com &        TV Movies \& Streaming &       1.453 &    0 &  1 &  0 &   1 &  1 &   1 &                  Netflix &  1 &         USA &  SCP \\
            dzen.ru &  Community \& Society - Other &       1.394 &    0 &  0 &  1 &   0 &  0 &   1 &                       VK &  1 &      Russia & Plat \\
microsoftonline.com &        Prog. \& Dev. Software &       1.239 &    1 &  0 &  0 &   1 &  0 &   1 &                Microsoft &  1 &         USA & SaaS \\
           bing.com &                Search Engines &       1.238 &    0 &  0 &  0 &   0 &  0 &   1 &                Microsoft &  1 &         USA & Plat \\
        samsung.com &          Consumer Electronics &       1.230 &    0 &  1 &  0 &   0 &  0 &   0 &                  Samsung &  1 & South Korea &  OCP \\
       bilibili.com &           Animation \& Comics &       1.200 &    0 &  - &  1 &   0 &  0 &   1 &                 Bilibili &  1 &       China & Plat \\
             vk.com &               Social Networks &       1.197 &    0 &  0 &  1 &   1 &  0 &   1 &                       VK &  1 &      Russia & Plat \\
          naver.com &                 News \& Media &       1.177 &    0 &  0 &  1 &   0 &  0 &   1 &                    Naver &  1 & South Korea & Plat \\
            mail.ru &                         Email &       1.170 &    1 &  1 &  1 &   1 &  0 &   1 &                       VK &  1 &      Russia & SaaS \\
        weather.com &                       Weather &       1.123 &    0 &  1 &  0 &   0 &  0 &   0 &                      IBM &  1 &         USA &  OCP \\
      pinterest.com &               Social Networks &       1.082 &    0 &  0 &  1 &   0 &  0 &   1 &                Pinterest &  1 &         USA & Plat \\
        discord.com &               Social Networks &       1.069 &    1 &  0 &  1 &   1 &  0 &   1 &                  Discord &  1 &         USA & Plat \\
          twitch.tv &                   Video Games &       1.038 &    0 &  0 &  1 &   0 &  0 &   1 &                   Amazon &  1 &         USA & Plat \\
     turbopages.org &                 News \& Media &       0.995 &    - &  - &  - &   - &  - &   - &                        - &  - &           - & Plat \\
            max.com &        TV Movies \& Streaming &       0.980 &    0 &  1 &  0 &   1 &  1 &   1 &   Warner Bros. Discovery &  1 &         USA &  SCP \\
      microsoft.com &        Prog. \& Dev. Software &       0.978 &    0 &  1 &  0 &   0 &  0 &   0 &                Microsoft &  1 &         USA &  OCP \\
            zoom.us &     Computers \& Tech - Other &       0.830 &    1 &  0 &  0 &   1 &  0 &   1 &                     Zoom &  1 &         USA & SaaS \\
     duckduckgo.com &                Search Engines &       0.806 &    0 &  0 &  0 &   0 &  0 &   1 &               DuckDuckGo &  1 &         USA & Plat \\
             qq.com &                 News \& Media &       0.791 &    0 &  - &  1 &   0 &  0 &   1 &                  Tencent &  1 &       China & Plat \\
      xhamster.desi &                         Adult &       0.776 &    0 &  0 &  1 &   0 &  0 &   1 &              Hammy Media &  1 &      Cyprus & Plat \\
         roblox.com &                   Video Games &       0.761 &    0 &  0 &  1 &   1 &  0 &   1 &                   Roblox &  1 &         USA & Plat \\
      stripchat.com &                         Adult &       0.737 &    0 &  0 &  1 &   0 &  0 &   1 &                StripChat &  1 &         USA & Plat \\
          quora.com & Dictionaries \& Encyclopedias &       0.729 &    0 &  0 &  1 &   1 &  0 &   1 &                    Quora &  1 &         USA & Plat \\
         fandom.com & Arts \& Entertainment - Other &       0.727 &    0 &  1 &  1 &   0 &  0 &   1 &                      TGP &  1 &         USA &  OCP \\
          globo.com &                 News \& Media &       0.723 &    0 &  1 &  0 &   0 &  0 &   1 &                    Globo &  1 &      Brazil &  OCP \\
     sharepoint.com &     Computers \& Tech - Other &       0.704 &    1 &  0 &  0 &   1 &  1 &   1 &                Microsoft &  1 &         USA & SaaS \\
               t.me &     Computers \& Tech - Other &       0.697 &    1 &  0 &  1 &   1 &  0 &   1 &                 Telegram &  1 &         UAE & Plat \\
           ebay.com &                   Marketplace &       0.688 &    0 &  0 &  1 &   0 &  0 &   1 &                     Ebay &  1 &         USA & Plat \\
            msn.com &                 News \& Media &       0.665 &    0 &  0 &  1 &   0 &  0 &   1 &                Microsoft &  1 &         USA & Plat \\
   news.yahoo.co.jp &                 News \& Media &       0.640 &    0 &  1 &  0 &   0 &  0 &   1 &                 SoftBank &  1 &       Japan &  OCP \\
            cnn.com &                 News \& Media &       0.622 &    0 &  1 &  0 &   0 &  0 &   1 &   Warner Bros. Discovery &  1 &         USA &  OCP \\
         indeed.com &            Jobs \& Employment &       0.612 &    0 &  0 &  1 &   1 &  0 &   1 &         Recruit Holdings &  1 &       Japan & Plat \\
        nytimes.com &                 News \& Media &       0.610 &    0 &  1 &  0 &   1 &  1 &   1 &       The New York Times &  1 &         USA &  SCP \\
        booking.com &       Accommodation \& Hotels &       0.583 &    0 &  0 &  1 &   0 &  0 &   1 &         Booking Holdings &  1 &         USA & Plat \\
      google.com.br &                Search Engines &       0.579 &    0 &  0 &  0 &   0 &  0 &   1 &                 Alphabet &  1 &         USA & Plat \\
\bottomrule
\end{tabular}

}
\caption{First half of the top 116 domains in terms of average monthly visits, along with our annotations and unsupervised cluster classification.}
\label{tab:annot-domains-1}
\end{table*}

\begin{table*}
\scalebox{0.88}{
  \begin{tabular}{llrllllllllll}
\toprule
             Domain &                  Industry &  V ($10^9$) & SaaS & CP & Pl & Log & Ch & Web &                       Owner & \$ &    Country &   Cl \\
\midrule
          bbc.co.uk &             News \& Media &       0.578 &    0 &  1 &  0 &   0 &  0 &   1 &                         BBC &  1 &         UK &  OCP \\
      spankbang.com &                     Adult &       0.574 &    0 &  0 &  1 &   0 &  0 &   1 &                   StripChat &  1 &        USA & Plat \\
       amazon.co.jp &               Marketplace &       0.570 &    0 &  1 &  1 &   0 &  0 &   0 &                      Amazon &  1 &        USA &  OCP \\
          aajtak.in &             News \& Media &       0.557 &    0 &  1 &  0 &   0 &  0 &   1 &          Living Media India &  1 &      India &  OCP \\
              ok.ru &           Social Networks &       0.551 &    0 &  0 &  1 &   1 &  0 &   1 &                          VK &  1 &     Russia & Plat \\
    accuweather.com &                   Weather &       0.547 &    0 &  1 &  0 &   0 &  0 &   0 &                 AccuWeather &  1 &        USA &  OCP \\
          zhihu.com &           Social Networks &       0.532 &    0 &  0 &  1 &   1 &  0 &   1 &                       Zhihu &  1 &      China & Plat \\
           espn.com &            Sports - Other &       0.528 &    0 &  1 &  0 &   1 &  1 &   1 &         Walt Disney Company &  1 &        USA &  SCP \\
            bbc.com &             News \& Media &       0.522 &    0 &  1 &  0 &   0 &  0 &   1 &                         BBC &  1 &         UK &  OCP \\
      rakuten.co.jp &               Marketplace &       0.507 &    0 &  - &  1 &   0 &  - &   - &                     Rakuten &  1 &      Japan & Plat \\
          canva.com &  Multimedia \& Web Design &       0.500 &    1 &  0 &  0 &   1 &  0 &   1 &                       Canva &  1 &  Australia & SaaS \\
          apple.com &      Consumer Electronics &       0.492 &    0 &  1 &  0 &   0 &  0 &   0 &                       Apple &  1 &        USA &  OCP \\
           imdb.com &    TV Movies \& Streaming &       0.488 &    0 &  0 &  1 &   0 &  0 &   1 &                      Amazon &  1 &        USA & Plat \\
              ya.ru &            Search Engines &       0.484 &    0 &  0 &  0 &   0 &  0 &   1 &                      Yandex &  1 &     Russia & Plat \\
        spotify.com &                     Music &       0.480 &    0 &  1 &  1 &   1 &  0 &   1 &                     Spotify &  1 &     Sweden &  OCP \\
      ssyoutube.com &            Search Engines &       0.475 &    1 &  0 &  0 &   0 &  0 &   1 &                           - &  - &          - & SaaS \\
    instructure.com &                 Education &       0.474 &    1 &  1 &  0 &   0 &  0 &   0 &                 Thoma Bravo &  1 &        USA &  OCP \\
         uol.com.br &             News \& Media &       0.472 &    0 &  1 &  0 &   0 &  0 &   1 &                 Grupo Folha &  1 &     Brazil &  OCP \\
       cricbuzz.com &            Fantasy Sports &       0.469 &    0 &  1 &  0 &   0 &  0 &   1 &             The Times Group &  1 &      India &  OCP \\
           etsy.com &               Marketplace &       0.466 &    0 &  0 &  1 &   0 &  0 &   1 &                        Etsy &  1 &        USA & Plat \\
      yiyouliao.com &    Prog. \& Dev. Software &       0.465 &    1 &  0 &  0 &   1 &  1 &   1 &         Youliang Technology &  1 &      China & SaaS \\
     aliexpress.com &               Marketplace &       0.454 &    0 &  0 &  1 &   0 &  0 &   1 &               Alibaba Group &  1 &      China & Plat \\
         paypal.com & Banking Credit \& Lending &       0.452 &    0 &  0 &  0 &   1 &  1 &   1 &                      Paypal &  1 &        USA & SaaS \\
          amazon.de &               Marketplace &       0.439 &    0 &  1 &  1 &   0 &  0 &   0 &                      Amazon &  1 &        USA &  OCP \\
     chaturbate.com &                     Adult &       0.427 &    0 &  0 &  1 &   0 &  0 &   1 &             Multi Media LLC &  1 &        USA & Plat \\
         github.com &    Prog. \& Dev. Software &       0.425 &    0 &  0 &  1 &   1 &  0 &   1 &                   Microsoft &  1 &        USA & Plat \\
    dailymail.co.uk &             News \& Media &       0.412 &    0 &  1 &  0 &   0 &  0 &   1 & Daily Mail \& General Trust &  1 &         UK &  OCP \\
          google.de &            Search Engines &       0.405 &    0 &  0 &  0 &   0 &  0 &   1 &                    Alphabet &  1 &        USA & Plat \\
            ozon.ru &               Marketplace &       0.403 &    0 &  1 &  1 &   0 &  0 &   0 &                        Ozon &  1 &     Russia &  OCP \\
         taobao.com &               Marketplace &       0.400 &    0 &  - &  1 &   1 &  - &   0 &               Alibaba Group &  1 &      China & Plat \\
        walmart.com &               Marketplace &       0.395 &    0 &  1 &  1 &   0 &  0 &   0 &               Walton family &  1 &        USA &  OCP \\
    news.google.com &             News \& Media &       0.392 &    0 &  0 &  0 &   0 &  0 &   1 &                    Alphabet &  1 &        USA & Plat \\
     wildberries.ru &               Marketplace &       0.391 &    0 &  1 &  1 &   0 &  0 &   0 &                 Wildberries &  1 &     Russia &  OCP \\
           avito.ru &               Classifieds &       0.378 &    0 &  0 &  1 &   0 &  0 &   1 &        Kismet Capital Group &  1 &     Russia & Plat \\
      miguvideo.com &    TV Movies \& Streaming &       0.365 &    0 &  - &  - &   - &  - &   1 &                           - &  - &          - & Plat \\
       amazon.co.uk &               Marketplace &       0.364 &    0 &  1 &  1 &   0 &  0 &   0 &                      Amazon &  1 &        USA &  OCP \\
    theguardian.com &             News \& Media &       0.352 &    0 &  1 &  0 &   0 &  0 &   1 &         Scott Trust Limited &  1 &         UK &  OCP \\
    tripmall.online &         Ecommerce - Other &       0.348 &    0 &  0 &  0 &   0 &  0 &   1 &                   12 Co Ltd &  1 &      Japan & Plat \\
          marca.com &            Sports - Other &       0.341 &    0 &  1 &  0 &   0 &  0 &   1 &         Cairo Communication &  1 &      Italy &  OCP \\
         pemsrv.com &                     Adult &       0.338 &    - &  - &  - &   - &  - &   - &                           - &  - &          - & Plat \\
      messenger.com &           Social Networks &       0.338 &    1 &  0 &  1 &   1 &  0 &   1 &                        Meta &  1 &        USA & Plat \\
          amazon.in &               Marketplace &       0.336 &    0 &  1 &  1 &   0 &  0 &   0 &                      Amazon &  1 &        USA &  OCP \\
         aniwave.to &       Animation \& Comics &       0.334 &    0 &  0 &  0 &   0 &  0 &   1 &                           - &  - &          - & Plat \\
          pixiv.net &     Visual Arts \& Design &       0.330 &    0 &  0 &  1 &   1 &  0 &   1 &                       Pixiv &  1 &      Japan & Plat \\
archiveofourown.org &       Books \& Literature &       0.327 &    0 &  0 &  1 &   0 &  0 &   1 &      Org. for Transf. Works &  0 &        USA & Plat \\
      pinduoduo.com &               Marketplace &       0.325 &    0 &  0 &  1 &   1 &  0 &   1 &                PDD Holdings &  1 &    Ireland & Plat \\
            aol.com & Computers \& Tech - Other &       0.324 &    0 &  0 &  0 &   0 &  0 &   1 &                      Apollo &  1 &        USA & Plat \\
           temu.com &               Marketplace &       0.321 &    0 &  0 &  1 &   1 &  0 &   1 &                PDD Holdings &  1 &    Ireland & Plat \\
         zillow.com &               Real Estate &       0.320 &    1 &  1 &  1 &   0 &  - &   0 &                      Zillow &  1 &        USA &  OCP \\
            line.me &           Social Networks &       0.314 &    1 &  0 &  1 &   1 &  0 &   1 &                    SoftBank &  1 &      Japan & Plat \\
     disneyplus.com &    TV Movies \& Streaming &       0.313 &    0 &  1 &  0 &   1 &  1 &   1 &         Walt Disney Company &  1 &        USA &  SCP \\
       google.co.jp &            Search Engines &       0.311 &    0 &  0 &  0 &   0 &  0 &   1 &                    Alphabet &  1 &        USA & Plat \\
       onlyfans.com &                     Adult &       0.308 &    0 &  0 &  1 &   1 &  1 &   1 &     Fenix International Ltd &  1 &         UK & Plat \\
      office365.com &    Prog. \& Dev. Software &       0.301 &    1 &  0 &  0 &   1 &  1 &   1 &                   Microsoft &  1 &        USA & SaaS \\
       telegram.org &           Social Networks &       0.296 &    1 &  0 &  1 &   1 &  0 &   1 &                    Telegram &  1 &        UAE & Plat \\
        foxnews.com &             News \& Media &       0.294 &    0 &  1 &  0 &   1 &  0 &   1 &              Murdoch family &  1 &        USA &  OCP \\
     tsyndicate.com &  Marketing \& Advertising &       0.290 &    0 &  0 &  0 &   0 &  0 &   1 &                           - &  - &          - & Plat \\
          seznam.cz &            Search Engines &       0.287 &    0 &  0 &  0 &   0 &  0 &   1 &                      Seznam &  1 & Czech Rep. & Plat \\
\bottomrule
\end{tabular}

}
\caption{Second half of the top 116 domains in terms of average monthly visits, along with our annotations and unsupervised cluster classification.}
\label{tab:annot-domains-2}
\end{table*}

\end{document}